\documentstyle[12pt]{article}

\begin{document}

\topmargin 0pt
\oddsidemargin 5mm
\def\bbox{{\,\lower0.9pt\vbox{\hrule \hbox{\vrule height 0.2 cm
\hskip 0.2 cm \vrule height 0.2 cm}\hrule}\,}}
\def\a{\alpha}
\def\b{\beta}
\def\g{\gamma}
\def\G{\Gamma}
\def\d{\delta}
\def\D{\Delta}
\def\e{\epsilon}
\def\ve{\varepsilon}
\def\z{\zeta}
\def\t{\theta}
\def\vt{\vartheta}
\def\r{\rho}
\def\vr{\varrho}
\def\k{\kappa}
\def\l{\lambda}
\def\L{\Lambda}
\def\m{\mu}
\def\n{\nu}
\def\o{\omega}
\def\O{\Omega}
\def\s{\sigma}
\def\vs{\varsigma}
\def\S{\Sigma}
\def\vphi{\varphi}
\def\av#1{\langle#1\rangle}
\def\pa{\partial}
\def\na{\nabla}
\def\hg{\hat g}
\def\un{\underline}
\def\ov{\overline}
\def\cF{{\cal F}}
\def\Hsl{H \hskip-8pt /}
\def\Fsl{F \hskip-6pt /}
\def\cFsl{\cF \hskip-5pt /}
\def\ksl{k \hskip-6pt /}
\def\pasl{\pa \hskip-6pt /}
\def\tr{{\rm tr}}
\def\tcF{{\tilde{\cal F}}}
\def\tg{{\tilde g}}
\def\cLNSI{{\cal L}^{\rm NS}_{\rm I}}
\def\bcLNSI{{\bar\cLNSI}}
\def\shalf{\frac{1}{2}}
\def\nn{\nonumber \\}
\def\ap#1{{\it Ann. Phys.} {\bf #1}}
\def\cmp#1{{\it Comm. Math. Phys.} {\bf #1}}
\def\cqg#1{{\it Class. Quantum Grav.} {\bf #1}}
\def\pl#1{{\it Phys. Lett.} {\bf B#1}}
\def\prl#1{{\it Phys. Rev. Lett.} {\bf #1}}
\def\prd#1{{\it Phys. Rev.} {\bf D#1}}
\def\prr#1{{\it Phys. Rev.} {\bf #1}}
\def\prb#1{{\it Phys. Rev.} {\bf B#1}}
\def\np#1{{\it Nucl. Phys.} {\bf B#1}}
\def\ncim#1{{\it Nuovo Cimento} {\bf #1}}
\def\jmp#1{{\it J. Math. Phys.} {\bf #1}}
\def\aam#1{{\it Adv. Appl. Math.} {\bf #1}}
\def\mpl#1{{\it Mod. Phys. Lett.} {\bf A#1}}
\def\ijmp#1{{\it Int. J. Mod. Phys.} {\bf A#1}}
\def\prep#1{{\it Phys. Rep.} {\bf #1C}}


\begin{titlepage}
\setcounter{page}{0}

\begin{flushright}
COLO-HEP-375 \\
hep-th/9609211 \\
September 1996
\end{flushright}

\vspace{5 mm}
\begin{center}
{\large Anomaly cancellation in M-theory }
\vspace{10 mm}

{\large S. P. de Alwis\footnote{e-mail:  
dealwis@gopika.colorado.edu}}\\
{\em Department of Physics, Box 390,
University of Colorado, Boulder, CO 80309}\\
\vspace{5 mm}
\end{center}
\vspace{10 mm}

\centerline{{\bf{Abstract}}}
We show the  complete cancellation of gauge and gravitational  
anomalies in the
M-theory of Horava and Witten using their boundary contribution, and  
a term
coming from the existence of two and five-branes. A factor of three  
discrepancy
noted in an earlier work is resolved.  We end with a comment on flux
quantization.

\end{titlepage}
\newpage
\renewcommand{\thefootnote}{\arabic{footnote}}
\setcounter{footnote}{0}

\setcounter{equation}{0}
In the M-theory of Horava and Witten \cite{hw} several consistency  
checks
associated with anomaly cancellation were verified. However one test  
which
involved a numerical coefficient
of a purely gravitational anomaly was not carried out.
In an earlier paper by the author \cite{sda}\footnote{In this paper  
it was also
found that the dimensionless ratio of gauge and gravitational  
couplings
determined by anomaly cancellation in \cite{hw} (see below) checks  
with a
result obtained from string dualtiy and D-brane methods, thus giving   
an
additional test of M-theory. } the coefficient of a certain M-theory
Green-Schwarz term \cite{dlm}   was determined, but  there  appeared  
to  be a
factor of three discrepancy with the expression (3.12) of \cite{hw}.  
In this
short note we review the anomaly cancellation argument and find that  
the
M-theory topological terms do indeed cancel both gauge and  
gravitational
anomalies. Finally we comment on flux quantization in M-theory in the  
light of
a recent paper by E. Witten\cite{ew}.

We work with the ``downstairs" version of the theory, i.e. on an 11-D   
manifold
$M=M_{10}\times S^1/Z_2$.
The topological term in the low energy effective action of M-theory  
is

\begin{equation}\label{a}
-{1\over \k^2}{1\over 6}\int_{M} C\wedge K\wedge K.
\end{equation}

In the above\footnote{It should be noted that (\ref{a}) has a factor  
two
compared to the usual term since we are working in the ``downstairs"  
version of
the theory with $M=M_{10}\times S^1/Z_2$ where the integral goes over  
half the
volume of the ``upstairs'" version where $M=M_{10}\times S^1$ and the  
fields
are $Z_2$ symmetric.} $C$ is the three form gauge field of 11D  
supergravity and
$K=dC$.\footnote{Our 11 D supergravity conventions and definitions  
are the same
as in \cite{dlm}. In particular we define a p-form gauge field as   
$A={1\over
p!}A_{I_1...I_p}dx^{I_1}...dx^{I_p}$. The field strength is then  
$F=dA$ or in
components $F_{I_1...I_{p+1}}=(p+1)\pa_{[I_1}A_{I_2...I_{p+1}]}$ with  
unit
strength anti-symmetrization. The comparison with
the notation of Horava and Witten is as follows. $K=\sqrt 2G,~  
C=\sqrt 2
C^{HW}$ where $C^{HW} = C^{HW}_{IJK}dx^Idx^Jdx^K ,  
G^{HW}=dC^{HW}={1\over
4!}G_{IJKL}dx^Idx^Jdx^Kdx^L$
are the three form gauge field   and field strength (called C and G  
in
\cite{hw}) as defined by Horava and Witten. Our indices I,J,K,L, run  
from 1 to
11 whilst indices A,B,C,D  run from 1 to 10.}
In the Horava-Witten theory the the manifold $M$ has a boundary   
which consists
of two
disconnected components on each of which $E_8$ gauge fields live, so  
that on
dimensional reduction to ten dimensions one  gets the low energy  
effective
action of the Heterotic $E_8\times E_8$ theory. On checking the  
supersymmetry
of the resulting theory it was found by Horava and  
Witten\footnote{See equation
(2.20) of the second paper of \cite{hw}.} that one needed to have, on  
each
component of the  boundary,

\begin{equation}\label{b}
K|_{\pa M} = {\k^2\over 2\l^2}\hat I_4.
\end{equation}
  In the above  $\l$ is the gauge field coupling,
\begin{equation}\label{c}
\hat I_4 ={1\over 2}\tr {\cal R}^2-\tr F^2,
\end{equation}
 $F$ is an $E_8$ gauge field strength, and $ \cal R$ is the curvature  
two form.
Now defining $Q_3 =\shalf \o_{3L}-\o_{3} $, where the two omegas  
correspond to
the Lorentz and gauge Chern-Simons forms, we have the standard  
descent
equations,\footnote{See for example reference \cite{gsw}, chapter  
13.}

\begin{equation}\label{d}
\hat I_4=dQ_3,~~\d Q_3=dQ^1_2.
\end{equation}
where $\d$ is a gauge and local Lorentz variation and $Q_2^1$ is a  
two-form
that is linear in the gauge parameter.

{}From (\ref{b}), the relation $K=dC$, and the first equation in  
(\ref{d}), we
have (up to an irrelevant exact form)

\begin{equation}\label{}
C|_{\pa M}={\k^2\over 2\l^2}Q_3.
\end{equation}
Hence from the second equation in ({\ref{d}})
\begin{equation}\label{e}
\d C|_{\pa M}={\k^2\over 2\l^2} dQ^1_2.
\end{equation}

Now clearly  we may extend this variation to the bulk by writing
\begin{equation}\label{varC}
\d C = d\L, ~~\L|_{\pa M}={\k^2\over 2\l^2} Q^1_2.
\end{equation}

 Hence we have from (\ref{a}),  and (\ref{varC})
\begin{eqnarray}\label{}
\d W&=&-{1\over\k^2}{1\over 6}\int_{M}d\L\wedge K\wedge K\nn
&=&-{1\over\k^2}{1\over 6}\left ({\k^2\over \l^2}\right  
)^3\shalf\int_{\pa
M}Q^1_2\wedge{\hat I_4^2\over 4},
\end{eqnarray}
where to get the second equality we have used Stokes' theorem,  
$dK=0$, and
(\ref{b}).

Now the boundary theory is anomalous and the variation of the quantum  
effective
action $\G$
is given by\footnote{The numerical coefficient in (\ref{g}) is fixed  
by
standard methods. See for example \cite{gsw} equation (13.3.41),  
(13.4.5) the
line before (13.5.6) and equations (13.5.5) and (13.5.8). The form of  
the
anomaly is given in \cite{hw}.}.
\begin{equation}\label{g}
\d\G =-{1\over 48(2\pi)^5}\int_{\pa M}Q_2^1\wedge\left ( -{\hat  
I^2_4\over
4}+X_8\right ),
\end{equation}
 where $X_8=-{1\over 8}\tr{\cal R}^4+{1\over 32}(\tr{\cal R}^2)^2$.
Cancellation of the $\hat I^2_4 $ part of the anomaly then determines

\begin{equation}\label{h}
\eta^{-1}\equiv{\k^4\over\l^6}={1\over 4(2\pi)^5}
\end{equation}
as in \cite{hw}.

Now as shown in \cite{dlm} the existence of two and five branes in  
the theory
implies that there
is an additional topological (Green-Schwarz) term in  
M-theory.\footnote{The
existence of this term may also be inferred from an earlier string  
theory
calculation \cite{vw}.}  This is given by,
\begin{equation}\label{i}
W_5=\left ({(2\pi)^2\over 2\k^2}\right )^{1/3}{1\over  
24(2\pi)^4}2\int _{
M}C\wedge X_8.
\end{equation}
The first factor in the equation above was obtained from the relation
$T_2=\left[ (2\pi )^2\over 2\k^2\right ]^{1/3}$, which was originally
determined using D-brane methods \cite{sda}, but after correcting a  
factor of
two in the  quantization formula of \cite{dlm} as discussed in the  
appendix to
\cite{sda},  it can also be fixed purely within  
M-theory.\footnote{There is an
extra
factor of 2 compared to equation (0.20) of  \cite{sda} because we are  
in the
``downstairs" theory - see footnote 2.}

Using equations (\ref{varC}), Stokes' theorem and $dX_8=0$ we have

\begin{eqnarray}\label{j}
\d W_5 &=& \left ({(2\pi)^2\over 2\k^2}\right )^{1/3}{1\over  
24(2\pi)^4}2\int
_{ M}d\L\wedge X_8\nn
&=&{1\over 48(2\pi)^5}\int_{\pa M}Q^1_2\wedge X_8.
\end{eqnarray}
In the last equation we have used the value of $\eta$  given in  
(\ref{h}). Thus
we have the complete cancellation of the anomalies in the  
Horava-Witten
M-theory,

\begin{equation}
\d W+\d W_5+\d\G=0.\noindent\\
\end{equation}

While the above was being written up, a paper by E. Witten appeared  
\cite{ew}
in which, {\it inter alia},
some issues of normalization in M-theory were discussed using index  
theory. To
conclude this note we would like to make some related comments.  
Equation
(\ref{b})
may be rewritten (after using \ref{h}) as

\begin{equation}\label{k}
{G\over 2\pi}|_{\pa M}\equiv\left [{(2\pi)^2\over2\k^2}\right  
]^{1/3}{K\over
2\pi}|_{\pa M} =w(V)-{\l \over 2},
\end{equation}
where $w=F\wedge F/(16\pi^2)$ has integer valued periods, being the  
second
Chern class  of the $E_8$ bundle, and
$\l\equiv {p_1/ 2}={{\cal R\wedge R}/ (16\pi^2)}$ which is half the  
Pontryagin
class $p_1$ of the tangent bundle, also has integer-valued periods  
for a spin
manifold so that in general as pointed out in \cite{ew}, ${G/ (2\pi  
)}$ has
half integer periods .  By considering a 4-cycle ($C$) in the bulk  
that is
homologous to one in the boundary ($C'$) this result was extended in  
\cite{ew}
to the statement
\begin{equation}\label{l}
\int_C \left ({G\over 2\pi}-{\l\over 2}\right )~\e~{\bf{\cal Z}}
\end{equation}
There is  an alternate way to get the  normalization of  $G/ (2\pi  
)$. This
follows from the fact that $K=dC$ and $C$ is the three form field  
coupling to
the 2-brane of M-theory. In earlier work
this was given as $T_2\int_C K/2\pi~\e~{\bf{\cal Z}}$\footnote{see  
for example
\cite{dlm} and the appendix of \cite{sda}.} but due to anomalies of  
fermionic
determinants in odd
dimensions, it was pointed out in \cite{ew} that this gets modified  
to

\begin{equation}\label{}
\int_C \left ({T_2K\over 2\pi}-{\l\over 2}\right )~\e~{\bf{\cal Z}}
\end{equation}
The consistency of the two normalizations is then a consequence of  
the relation
$T_2=\left[ (2\pi )^2\over 2\k^2\right ]^{1/3}$ derived in  
\cite{sda}. Note
that this is a check on the consistency of M-theory that is  
independent of the
check coming from the pure gravity anomaly cancellation.\footnote{As  
noted in
\cite{sda} (equation (8), (11) and the discussion after the latter  
equation),
the calculation of $T_2$
can be done in a fashion that is completely independent of M-theory
quantization conditions.}

\noindent{\bf Acknowledgments}\\
I wish to thank E. Witten for the suggestion to check M-theory  
consistency
conditions, and S. Chaudhuri and  P. Horava for discussions. This  
work is
partially supported by the
Department of Energy contract No. DE-FG02-91-ER-40672.


\end{document}